# 量子隨機數產生器應用於 NIST 後量子密碼學標準演算法

Abel C. H. Chen
*Information & Communications Security Laboratory,
Chunghwa Telecom Laboratories*
Taoyuan, Taiwan
Email address: chchen.scholar@gmail.com, ORCID: 0000-0003-3628-3033

*摘要*—近年來隨著量子計算技術的成熟，將可能對 RSA 密碼學和橢圓曲線密碼學產生安全上的威脅。因此，國家標準暨技術研究院(National Institute of Standards and Technology, NIST) 在 2024 年 8 月陸續制定了模晶格金鑰封裝機制(Module-Lattice-Based Key-Encapsulation Mechanism, ML-KEM)、模晶格數位簽章演算法(Module-Lattice-Based Digital Signature Algorithm, ML-DSA)、無狀態雜湊數位簽章演算法(Stateless Hash-Based Digital Signature Algorithm, SLH-DSA)等聯邦資訊處理標準(Federal Information Processing Standards, FIPS)。然而，雖然這些後量子密碼學標準演算法具備抵抗量子計算攻擊的能力，但在特殊應用情境下可能不夠安全。有鑑於此，本研究提出基於量子隨機數產生器的後量子密碼學演算法，可以通過量子計算產生隨機數，並且在量子隨機數產生器基礎上產製金鑰對、產製金鑰封裝，以及產生數位簽章。本研究提出一個泛化(general)的量子隨機數產生器架構，並且設計 6 種量子隨機數產生器，分別運用 NIST SP 800-90B 所述驗證方法驗證符合隨機位元熵(entropy)驗證條件和隨機位元獨立且同分布(Independent and Identically Distributed, IID)驗證條件。在實驗中，本研究亦驗證 6 種量子隨機數產生器的計算時間，以及驗證 QRNG-Based ML-KEM、QRNG-Based ML-DSA、QRNG-Based SLH-DSA 的計算時間，可供未來部署參考。

*關鍵字*—量子隨機數產生器、真隨機數、模晶格金鑰封裝機制、模晶格數位簽章演算法、無狀態雜湊數位簽章演算法

## I. 前言

為了因應量子計算對 RSA 密碼學和橢圓曲線密碼學造成的安全威脅[1]，國家標準暨技術研究院(National Institute of Standards and Technology, NIST)從 2016 年開始徵求後量子密碼學演算法，為制定後量子密碼學標準做出了卓越的貢獻。其中，在 2024 年 8 月分別制定了模晶格金鑰封裝機制(Module-Lattice-Based Key-Encapsulation Mechanism, ML-KEM)[2]、模晶格數位簽章(Module-Lattice-Based Digital Signature Algorithm, ML-DSA)[3]、無狀態雜湊數位簽章(Stateless Hash-Based Digital Signature Algorithm, SLH-DSA)[4]等聯邦資訊處理標準(Federal Information Processing Standards, FIPS)，並且未來預計會制定 FN-DSA (FFT (Fast-Fourier Transform) over NTRU (Nth-degree Truncated polynomial Ring Unit)-Lattice-Based DSA)[5]和 HQC (Hamming Quasi-Cyclic)[6]相關標準演算法。除此之外，國家標準暨技術研究院在 2024 年 11 月出版了後量子密碼學標準遷移草案[7]，明確了遷移時程，反應出引入後量子密碼學是近期重要的發展方向。

國家標準暨技術研究院負責後量子密碼學專案的重要成員 Yi-Kai Liu 和 Dustin Moody 於 2024 年在《Physical Review Applied》發表一篇很棒的論文「Post-quantum cryptography and the quantum future of cybersecurity」整理了後量子密碼學和量子計算的近期發展，並且討論其各自的優勢和限制[8]。除此之外，該論文也指出後量子密碼學和量子計算可以有互補的作法，進行後量子密碼學和量子計算的結合將有可能提升安全性。有鑑於此，本研究提出一個泛化(general)的量子隨機數產生器架構，並且建構 6 種量子隨機數產生器，並且設計 QRNG-Based ML-KEM、QRNG-Based ML-DSA、QRNG-Based SLH-DSA，運用量子隨機數產生器產製金鑰對、產製金鑰封裝、以及產製數位簽章。

本研究的主要貢獻條列如下：

- 本研究提出一個泛化(general)的量子隨機數產生器架構，並且運用 6 種不同的邏輯閘組合產生均勻疊加態，再量測得到隨機位元值。

- 本研究參考 NIST SP 800-90B 所述驗證方法[9]，驗證本研究提出的 6 種量子隨機數產生器，符合隨機位元熵(entropy)驗證條件和隨機位元獨立且同分布(Independent and Identically Distributed, IID)驗證條件。

- 本研究將量子隨機數產生器應用於金鑰封裝機制，以 ML-KEM 為例，應用在產製金鑰對和產製金鑰封裝。

- 本研究將量子隨機數產生器應用於數位簽章，以 ML-DSA 和 SLH-DSA 為例，應用在產製金鑰對和產製數位簽章。

本文主要分為 7 個章節。第 II 節說明本研究設計的量子隨機數產生器架構及 6 種量子隨機數產生器。第 III 節~第 V 節分別描述 QRNG-Based ML-KEM、QRNG-Based ML-DSA、QRNG-Based SLH-DSA 的細部設計。第 VI 節驗證本研究設計的量子隨機數產生器其隨機性及計算效率，以及驗證 QRNG-Based ML-KEM、QRNG-Based ML-DSA、QRNG-Based SLH-DSA 的計算時間。最後，第 VII 節總結本研究貢獻，並且討論研究限制及未來研究方向。

## II. 量子隨機數產生器

本節將深入描述本研究設計的量子隨機數產生器。其中，在第 II.A 小節提出一個泛化(general)的量子隨機數產生器架構，之後在第 II.B 小節~第 II.G 小節分別採用不同的量子邏輯閘來產生均勻疊加態，建構 6 種量子隨機數產生器。

### A. 量子隨機數產生器架構

本研究設計的量子隨機數產生器核心精神在於運用均勻疊加態產生器(包含 1 個或多個量子邏輯閘)來為每個量子位元產生均勻疊加態，之後再對均勻疊加態情況下的量子位元進行量測得到經典位元的隨機位元值。其中，由於均勻疊加態情況下的量子位元量測到 0 或量測到 1 的機率各為 0.5，所以可以達到較高的熵。

本研究設計的量子隨機數產生器架構主要為一個 $c$ 個量子位元的量子電路，並且每個量子位元都會經過均勻疊加態產生器操作，之後再對均勻疊加態的量子位元進行量測，架構圖如圖 1 所示。其中，當欲產生 $L$ 長度的隨機位元時，則可以對量子電路循環操作 $\left\lfloor \frac{L}{c} \right\rfloor$ 次即可。例如，如果想最小化量子計算硬體成本，可以只建構 1 個量子位元(即 $c = 1$)的量子電路；當欲產製 256 位元長度(即 $L = 256$)的隨機數時，則需要對量子電路循環操作 256 次。

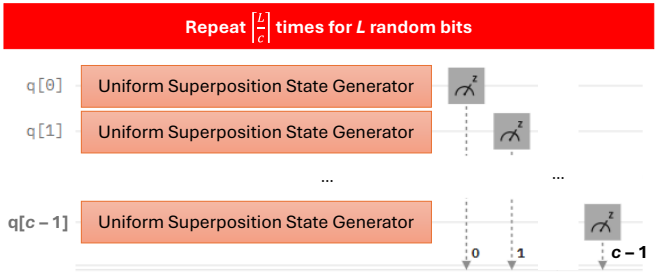

Fig. 1. 量子隨機數產生器架構

第 II.B 小節~第 II.G 小節將分別運用 Hadamard-Gate (H-Gate)、Square-root of X-Gate (SX-Gate)、Rotation-X-Gate (RX-Gate)、Rotation-Y-Gate (RY-Gate)、Phase-Gate (P-Gate)、以及 Universal Single-Qubit-Gate (U-Gate)[10]的組合來建構量子隨機數產生器。其中，為簡化說明，本研究以 1 個量子位元的量子電路為例，並且假設量子位元的初始值為 $|0\rangle$，也就是可以用向量 $\begin{bmatrix}1\\0\end{bmatrix}$ 表示其量子態。

### B. 方法(1)：基於 H-Gate 量子隨機數產生器

這個小節是最常見的方法，通過操作 H-Gate (Matrix Representation 如公式(1)所示)來產生均勻疊加態，如公式(2)所示。因此，量測到 0 的機率是 $\left|\frac{1}{\sqrt{2}}\right|^2 = 0.5$，量測到 1 的機率是 $\left|\frac{1}{\sqrt{2}}\right|^2 = 0.5$。

$$H = \frac{1}{\sqrt{2}}\begin{bmatrix}1 & 1\\1 & -1\end{bmatrix} \quad (1)$$

$$H|0\rangle = \frac{1}{\sqrt{2}}\begin{bmatrix}1 & 1\\1 & -1\end{bmatrix}\begin{bmatrix}1\\0\end{bmatrix} = \frac{1}{\sqrt{2}}\begin{bmatrix}1\\1\end{bmatrix} \quad (2)$$

### C. 方法(2)：基於 SX-Gate 量子隨機數產生器

這個小節是通過操作 SX-Gate (Matrix Representation 如公式(3)所示)來產生均勻疊加態，如公式(4)所示。因此，量測到 0 的機率是 $\left|\frac{1+i}{2}\right|^2 = 0.5$，量測到 1 的機率是 $\left|\frac{1-i}{2}\right|^2 = 0.5$。

$$SX = \frac{1}{2}\begin{bmatrix}1+i & 1-i\\1-i & 1+i\end{bmatrix} \quad (3)$$

$$SX|0\rangle = \frac{1}{2}\begin{bmatrix}1+i & 1-i\\1-i & 1+i\end{bmatrix}\begin{bmatrix}1\\0\end{bmatrix} = \frac{1}{2}\begin{bmatrix}1+i\\1-i\end{bmatrix} \quad (4)$$

### D. 方法(3)：基於 RX-Gate 量子隨機數產生器

這個小節是通過操作 RX-Gate (Matrix Representation 如公式(5)所示)對 X 軸旋轉 $\frac{\pi}{2}$ (即 $\theta = \frac{\pi}{2}$)來產生均勻疊加態，如公式(6)所示。因此，量測到 0 的機率是 $\left|\frac{1}{\sqrt{2}}\right|^2 = 0.5$，量測到 1 的機率是 $\left|-\frac{i}{\sqrt{2}}\right|^2 = 0.5$。

$$RX(\theta) = \begin{bmatrix}\cos\left(\frac{\theta}{2}\right) & -i\sin\left(\frac{\theta}{2}\right)\\-i\sin\left(\frac{\theta}{2}\right) & \cos\left(\frac{\theta}{2}\right)\end{bmatrix} \quad (5)$$

$$RX\left(\frac{\pi}{2}\right)|0\rangle = \begin{bmatrix}\cos\left(\frac{\pi}{4}\right) & -i\sin\left(\frac{\pi}{4}\right)\\-i\sin\left(\frac{\pi}{4}\right) & \cos\left(\frac{\pi}{4}\right)\end{bmatrix}\begin{bmatrix}1\\0\end{bmatrix}$$

$$= \begin{bmatrix}\cos\left(\frac{\pi}{4}\right)\\-i\sin\left(\frac{\pi}{4}\right)\end{bmatrix} = \begin{bmatrix}\frac{1}{\sqrt{2}}\\-\frac{i}{\sqrt{2}}\end{bmatrix} \quad (6)$$

### E. 方法(4)：基於 RY-Gate 量子隨機數產生器

這個小節是通過操作 RYGate (Matrix Representation 如公式(7)所示)對 Y 軸旋轉 $\frac{\pi}{2}$ (即 $\theta = \frac{\pi}{2}$)來產生均勻疊加態，如公式(8)所示。因此，量測到 0 的機率是 $\left|\frac{1}{\sqrt{2}}\right|^2 = 0.5$，量測到 1 的機率是 $\left|\frac{1}{\sqrt{2}}\right|^2 = 0.5$。

$$RY(\theta) = \begin{bmatrix}\cos\left(\frac{\theta}{2}\right) & -\sin\left(\frac{\theta}{2}\right)\\\sin\left(\frac{\theta}{2}\right) & \cos\left(\frac{\theta}{2}\right)\end{bmatrix} \quad (7)$$

$$RY\left(\frac{\pi}{2}\right)|0\rangle = \begin{bmatrix}\cos\left(\frac{\pi}{4}\right) & -\sin\left(\frac{\pi}{4}\right)\\\sin\left(\frac{\pi}{4}\right) & \cos\left(\frac{\pi}{4}\right)\end{bmatrix}\begin{bmatrix}1\\0\end{bmatrix}$$

$$= \begin{bmatrix}\cos\left(\frac{\pi}{4}\right)\\\sin\left(\frac{\pi}{4}\right)\end{bmatrix} = \begin{bmatrix}\frac{1}{\sqrt{2}}\\\frac{1}{\sqrt{2}}\end{bmatrix} \quad (8)$$

### F. 方法(5)：基於 P-Gate & H-Gate 量子隨機數產生器

這個小節是通過操作 P-Gate (Matrix Representation 如公式(9)所示)旋轉 $\frac{\pi}{2}$ (即 $\theta = \frac{\pi}{2}$)和 H-Gate 來產生均勻疊加態，

如公式(10)所示。因此，量測到 0 的機率是 $\left|\frac{1}{\sqrt{2}}\right|^2 = 0.5$，量測到 1 的機率是 $\left|\frac{1}{\sqrt{2}}\right|^2 = 0.5$。

$$P(\theta) = \begin{bmatrix} 1 & 0 \\ 0 & e^{i\theta} \end{bmatrix} \quad (9)$$

$$HP(\theta)|0\rangle = \frac{1}{\sqrt{2}}\begin{bmatrix} 1 & 1 \\ 1 & -1 \end{bmatrix}\begin{bmatrix} 1 & 0 \\ 0 & e^{i\theta} \end{bmatrix}\begin{bmatrix} 1 \\ 0 \end{bmatrix} = \frac{1}{\sqrt{2}}\begin{bmatrix} 1 \\ 1 \end{bmatrix} \quad (10)$$

### G. 方法(6)：基於 U-Gate 量子隨機數產生器

這個小節是通過操作 U-Gate (Matrix Representation 如公式(11)所示)對先 Z 軸旋轉 $\frac{\pi}{2}$、再對 Y 軸旋轉 $\frac{\pi}{2}$、以及對 Z 軸旋轉 $\frac{\pi}{2}$ 來產生均勻疊加態，如公式(12)所示。因此，量測到 0 的機率是 $\left|\frac{1}{\sqrt{2}}\right|^2 = 0.5$，量測到 1 的機率是 $\left|\frac{i}{\sqrt{2}}\right|^2 = 0.5$。

$$U\left(\frac{\pi}{2},\frac{\pi}{2},\frac{\pi}{2}\right) = \begin{bmatrix} \cos\left(\frac{\pi}{4}\right) & -e^{i\frac{\pi}{2}}\sin\left(\frac{\pi}{4}\right) \\ e^{i\frac{\pi}{2}}\sin\left(\frac{\pi}{4}\right) & e^{i\pi}\cos\left(\frac{\pi}{4}\right) \end{bmatrix} \quad (11)$$

$$U\left(\frac{\pi}{2},\frac{\pi}{2},\frac{\pi}{2}\right)|0\rangle = \begin{bmatrix} \cos\left(\frac{\pi}{4}\right) & -e^{i\frac{\pi}{2}}\sin\left(\frac{\pi}{4}\right) \\ e^{i\frac{\pi}{2}}\sin\left(\frac{\pi}{4}\right) & e^{i\pi}\cos\left(\frac{\pi}{4}\right) \end{bmatrix}\begin{bmatrix} 1 \\ 0 \end{bmatrix}$$
$$= \begin{bmatrix} \cos\left(\frac{\pi}{4}\right) \\ e^{i\frac{\pi}{2}}\sin\left(\frac{\pi}{4}\right) \end{bmatrix} = \begin{bmatrix} \frac{1}{\sqrt{2}} \\ \frac{i}{\sqrt{2}} \end{bmatrix} \quad (12)$$

### H. 基於量子糾纏態的量子隨機數產生器

論文「Post-quantum cryptography and the quantum future of cybersecurity」也有提及可以通過量子金鑰分配(Quantum Key Distribution, QKD)的方式來產製隨機數[8]。有鑑於此，本研究可以修改前述方法，結合量子金鑰分配(以 E91 協定[11]為例)，由兩個硬體安全模組(Hardware Security Module, HSM)各自分別持有處於量子糾纏態的量子位元。如圖 2 所示，量子位元 q[0]和 q[1]處於量子糾纏態，HSM 1 持有量子位元 q[0]，HSM 2 持有量子位元 q[1]。HSM 1 和 HSM 2 可以各自隨機選擇控制訊號(圖 2 的例子為 HSM 1 和 HSM 2 剛好都選擇 P-Gate & H-Gate 作為控制訊號)後進行量測，並且循環多次後，最後再比對控制訊號序列決定要選擇哪些量測位元值作為協商後的值。由於本研究主要採用第 II.A 小節~第 II.G 小節的方式來建構量子隨機數產生器，所以未對基於量子糾纏態的量子隨機數產生器深入著墨。

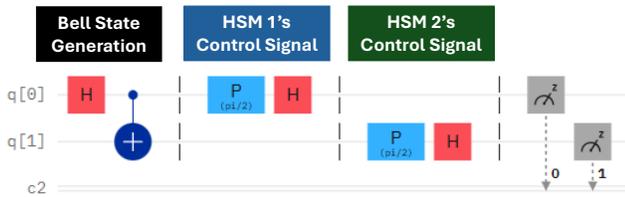

Fig. 2. 基於量子糾纏態的量子隨機數產生器

### III. 量子隨機數產生器應用於模晶格金鑰封裝機制

FIPS 203 定義了模晶格金鑰封裝機制的相關參數和具體演算法。本節主要著重在討論哪些設計可以修改為採用量子隨機數產生器來提升安全性。

#### A. 產製金鑰對

模晶格金鑰封裝機制的產製金鑰對演算法主要定義在 FIPS 203 的 Algorithm 19 [2]。其中，主要將產製兩個各 32 bytes 的隨機數，分別為 $d$ 和 $z$，並且將代入 ML-KEM.KeyGen_internal($d$, $z$)計算後得到公鑰 $ek$ 和私鑰 $dk$ [2]。有鑑於此，本研究修改演算法，由量子隨機數產生器產製兩個 256 bits 的隨機數作為 $d$ 和 $z$，修改後演算法如本研究的 **Algorithm 1**。

| **Algorithm 1** ML-KEM.KeyGen() |
|---|
| **Output**: encapsulation key $ek$. |
| **Output**: decapsulation key $dk$. |
| 1: $d \leftarrow$ **256 random bits from QRNG** |
| 2: $z \leftarrow$ **256 random bits from QRNG** |
| 3: **if** $d$ == NULL **or** $z$ == NULL **then** |
| 4:     **return** $\perp$ |
| 5: **end if** |
| 6: ($ek$, $dk$) $\leftarrow$ ML-KEM.KeyGen_internal($d$, $z$) |
| 7: **return** ($ek$, $dk$) |

#### B. 產製金鑰封裝

為了達到在選擇密文攻擊下的不可區分性(Indistinguishability under Chosen-Ciphertext Attack, IND-CCA)安全等級，所以每次產製密文時也都會搭配不同的隨機數。因此，在模晶格金鑰封裝機制的產製金鑰封裝演算法主要定義在 FIPS 203 的 Algorithm 20 [2]。其中，運用公鑰 $ek$ 和隨機數 $m$ 來代入 ML-KEM.Encaps_internal($ek$, $m$)產製共享秘密金鑰(shared secret key) $K$ 和偽隨機數 $r$，再通過 K-PKE.Encrypt($ek$, $m$, $r$)得到密文 $c$ [2]。有鑑於此，本研究修改演算法，由量子隨機數產生器產製一個 256 bits 的隨機數作為 $m$，修改後演算法如本研究的 **Algorithm 2**。

| **Algorithm 2** ML-KEM.Encaps($ek$) |
|---|
| **Input**: encapsulation key $ek$. |
| **Output**: shared secret key $K$. |
| **Output**: ciphertext $c$. |
| 1: $m \leftarrow$ **256 random bits from QRNG** |
| 2: **if** $m$ == NULL **then** |
| 3:     **return** $\perp$ |
| 4: **end if** |
| 5: ($K$, $c$) $\leftarrow$ ML-KEM.Encaps_internal($ek$, $m$) |
| 6: **return** ($K$, $c$) |

## C. 產製多個隨機數需求

當有需要產製多個隨機數時，可以呼叫一次或多次量子隨機數產生器。以模晶格金鑰封裝機制的產製金鑰對演算法為例，需要產製兩個各 32 bytes 的隨機數，可以呼叫量子隨機數產生器兩次各產製 32 bytes 的隨機數。然而，也可以僅呼叫量子隨機數產生器一次產製 64 bytes 的隨機數，取前 32 bytes 作為 $d$，取後 32 bytes 作為 $z$。當量子隨機數產生器產製的隨機位元符合獨立且同分布時，呼叫一次和呼叫多次是等價的。

## IV. 量子隨機數產生器應用於模晶格數位簽章演算法

FIPS 204 定義了模晶格數位簽章演算法的相關參數和具體演算法。本節主要著重在討論哪些設計可以修改為採用量子隨機數產生器來提升安全性。

### A. 產製金鑰對

模晶格數位簽章演算法的產製金鑰對演算法主要定義在 FIPS 204 的 Algorithm 1 [3]。其中，主要將產製一個 32 bytes 的隨機數 $\xi$，並且將代入 ML-DSA.KeyGen_internal($\xi$) 計算後得到公鑰 $pk$ 和私鑰 $sk$ [3]。有鑑於此，本研究修改演算法，由量子隨機數產生器產製一個 256 bits 的隨機數作為 $\xi$，修改後演算法如本研究的 **Algorithm 3**。

---

**Algorithm 3** ML-DSA.KeyGen()

**Output**: public key $pk$.

**Output**: private key $sk$.

1: $\xi \leftarrow$ **256 random bits from QRNG**

2: **if** $\xi ==$ NULL **then**

3:     **return** $\perp$

4: **end if**

5: **return** ML-DSA.KeyGen_internal($\xi$)

---

### B. 產製數位簽章

為了達到在選擇訊息攻擊下的存在性不可偽造性 (Existential Unforgeability under Chosen Message Attack, EUF-CMA)安全等級，所以每次產製簽章時也都會搭配不同的隨機數。因此，在模晶格數位簽章演算法的產製簽章演算法主要定義在 FIPS 204 的 Algorithm 2 [3]。其中，運用私鑰 $sk$、待簽章訊息 $M'$、以及一個 32 bytes 的隨機數 $rnd$ 來代入 ML-DSA.Sign_internal($sk$, $M'$, $rnd$)產製簽章 $\sigma$ [3]。有鑑於此，本研究修改演算法，由量子隨機數產生器產製一個 256 bits 的隨機數作為 $rnd$，修改後演算法如本研究的 **Algorithm 4**。

### C. 產製預雜湊數位簽章

FIPS 204 的 Algorithm 4 也定義了預雜湊模晶格數位簽章演算法(Pre-Hash ML-DSA)，通過代入指定的預雜湊函數 $PH$，對訊息 $M$ 進行雜湊計算後得到 $PH_M$，再根據 OID 和 $PH_M$ 得到待簽章訊息 $M'$，其他計算與 FIPS 204 的 Algorithm 2 一致[3]。有鑑於此，本研究修改演算法，由量子隨機數產生器產製一個 256 bits 的隨機數作為 $rnd$，修改後演算法如本研究的 **Algorithm 5**。

---

**Algorithm 4** ML-DSA.Sign($sk$, $M$, $ctx$)

**Input**: private key $sk$, message $M$, context string $ctx$.

**Output**: signature $\sigma$.

1: **if** $|ctx| > 255$ **then**

2:     **return** $\perp$

3: **end if**

4:

5: $rnd \leftarrow$ **256 random bits from QRNG**

6: **if** $rnd ==$ NULL **then**

7:     **return** $\perp$

8: **end if**

9:

10: $M' \leftarrow$ BytesToBits(IntegerToBytes(0, 1) ||
       IntegerToBytes($|ctx|$, 1) || $ctx$) || $M$

11: $\sigma \leftarrow$ ML-DSA.Sign_internal($sk$, $M'$, $rnd$)

12: **return** $\sigma$

---

**Algorithm 5** HashML-DSA.Sign($sk$, $M$, $ctx$, PH)

**Input**: private key $sk$, message $M$, context string $ctx$,
       pre-hash function $PH$.

**Output**: signature $\sigma$.

1: **if** $|ctx| > 255$ **then**

2:     **return** $\perp$

3: **end if**

4:

5: $rnd \leftarrow$ **256 random bits from QRNG**

6: **if** $rnd ==$ NULL **then**

7:     **return** $\perp$

8: **end if**

9: execute the flows from **Line 9** to **Line 22** in
   **Algorithm 4** of FIPS 204 to get OID and generate the hashed message $PH_M$ based on PH.

…

23: $M' \leftarrow$ BytesToBits(IntegerToBytes(1, 1) ||
        IntegerToBytes($|ctx|$, 1) || $ctx$ || OID || $PH_M$)

24: $\sigma \leftarrow$ ML-DSA.Sign_internal($sk$, $M'$, $rnd$)

25: **return** $\sigma$

---

## V. 量子隨機數產生器應用於無狀態雜湊數位簽章演算法

FIPS 205 定義了無狀態雜湊數位簽章演算法的相關參數和具體演算法。本節主要著重在討論哪些設計可以修改為採用量子隨機數產生器來提升安全性。

### A. 產製金鑰對

無狀態雜湊數位簽章演算法的產製金鑰對演算法主要定義在 FIPS 205 的 Algorithm 21 [4]。其中，主要將產製三個 $n$ bytes 的隨機數，分別是 SK.seed、SK.prf、PK.seed，並且將代入 slh_keygen_internal(SK.seed, SK.prf, PK.seed)計算後得到公鑰 PK 和私鑰 SK [4]。有鑑於此，本研究修改演算法，由量子隨機數產生器產製三個 $8n$ bits 的隨機數分別作為 SK.seed、SK.prf、PK.seed，修改後演算法如本研究的 **Algorithm 6**。

---

**Algorithm 6** slh_keygen()

**Output**: SLH-DSA key pair (SK, PK).

1: SK.seed ← **$8n$ random bits from QRNG**

2: SK.prf ← **$8n$ random bits from QRNG**

3: PK.seed ← **$8n$ random bits from QRNG**

4: **if** SK.seed == NULL **or** SK.prf == NULL **or** PK.seed == NULL **then**

5:　　**return** ⊥

6: **end if**

7: **return** slh_keygen_internal(SK.seed, SK.prf, PK.seed)

---

在無狀態雜湊數位簽章演算法中具有不同的參數組合，包含 SLH-DSA-SHA2-128s、SLH-DSA-SHAKE-128f 等，其對應的 $n$ 值不同，如表 I 所示[4]。例如，如果採用 SLH-DSA-SHAKE-128f，則對應的 $n$ 值為 16，所以產製金鑰時需要三個 128 bits 的隨機數。

TABLE I. SLH-DSA 參數 $n$ 值

|  | $n$ | $8n$ | Security Level |
|---|---|---|---|
| SLH-DSA-SHA2-128s<br>SLH-DSA-SHAKE-128s | 16 | 128 | 1 |
| SLH-DSA-SHA2-128f<br>SLH-DSA-SHAKE-128f | 16 | 128 | 1 |
| SLH-DSA-SHA2-192s<br>SLH-DSA-SHAKE-192s | 24 | 192 | 3 |
| SLH-DSA-SHA2-192f<br>SLH-DSA-SHAKE-192f | 24 | 192 | 3 |
| SLH-DSA-SHA2-256s<br>SLH-DSA-SHAKE-256s | 32 | 256 | 5 |
| SLH-DSA-SHA2-256f<br>SLH-DSA-SHAKE-256f | 32 | 256 | 5 |

如第 III.C 節所述，當量子隨機數產生器產製的隨機位元符合獨立且同分布時，呼叫一次和呼叫多次是等價的。因此，產製 SLH-DSA-SHAKE-128f 的金鑰對時，可以呼叫量子隨機數產生器一次，產製 384 bits 的隨機數。其中，第 1 個 bit~第 128 個 bit 作為 SK.seed、第 129 個 bit~第 256 個 bit 作為 SK.prf、以及第 257 個 bit~第 384 個 bit 作為 PK.seed。

### B. 產製數位簽章

為了達到在選擇訊息攻擊下的存在性不可偽造性(EUF-CMA)安全等級，所以每次產製簽章時也都會搭配不同的隨機數。因此，在無狀態雜湊數位簽章演算法的產製簽章演算法主要定義在 FIPS 205 的 Algorithm 22 [4]。其中，運用待簽章訊息 $M'$、私鑰 SK、以及一個 $n$ bytes 的隨機數 *addrnd* 來代入 slh_sign_internal($M'$, SK, *addrnd*)產製簽章 SIG [4]。有鑑於此，本研究修改演算法，由量子隨機數產生器產製一個 $8n$ bits 的隨機數作為 *addrnd*，修改後演算法如本研究的 **Algorithm 7**。例如，產製 SLH-DSA-SHAKE-128f 的數位簽章時，由量子隨機數產生器產製一個 128 bits 的隨機數作為 *addrnd*。

---

**Algorithm 7** slh_sign(*M*, *ctx*, SK)

**Input**: message *M*, context string *ctx*, private key SK.

**Output**: SLH-DSA signature SIG.

1: **if** |*ctx*| > 255 **then**

2:　　**return** ⊥

3: **end if**

4: *addrnd* ← **$8n$ random bits from QRNG**

5: **if** *addrnd* == NULL **then**

6:　　**return** ⊥

7: **end if**

8: $M'$ ← toByte(0, 1) || toByte(|*ctx*|, 1) || *ctx* || *M*

9: SIG ← slh_sign_internal($M'$, SK, *addrnd*)

10: **return** SIG

---

### C. 產製預雜湊數位簽章

FIPS 205 的 Algorithm 23 也定義了預雜湊無狀態雜湊數位簽章演算法(Pre-Hash SLH-DSA)，通過代入指定的預雜湊函數 PH，對訊息 $M$ 進行雜湊計算後得到 $PH_M$，再根據 OID 和 $PH_M$ 得到待簽章訊息 $M'$，其他計算與 FIPS 205 的 Algorithm 22 一致[4]。有鑑於此，本研究修改演算法，由量子隨機數產生器產製一個 $8n$ bits 的隨機數作為 *addrnd*，修改後演算法如本研究的 **Algorithm 8**。例如，產製 SLH-DSA-SHAKE-128f 的數位簽章時，由量子隨機數產生器產製一個 128 bits 的隨機數作為 *addrnd*。

## VI. 實驗結果與討論

本節將對本研究設計的 6 種量子隨機數產生器、QRNG-Based ML-KEM、QRNG-Based ML-DSA、QRNG-Based SLH-DSA 進行效能驗證。首先，第 VI.A 節說明實驗環境和相關設置。第 IV.B 節和第 IV.C 節驗證量子隨機數產生器產製的隨機位元，以及第 IV.D 節分析 6 種量子隨機

數產生器的計算時間。最後，第 IV.E 節討論量子隨機數產生器應用於 NIST 後量子密碼學標準演算法的計算時間。

**Algorithm 8** hash_slh_sign(*M*, *ctx*, PH, SK)

**Input**: message *M*, context string *ctx*, pre-hash function *PH*, private key SK.

**Output**: SLH-DSA signature SIG.

1: **if** |*ctx*| > 255 **then**

2:　　**return** ⊥

3: **end if**

4: *addrnd* ← **8*n* random bits from QRNG**

5: **if** *addrnd* == NULL **then**

6:　　**return** ⊥

7: **end if**

8: execute the flows from **Line 8** to **Line 23** in

　**Algorithm 23** of FIPS 205 to get OID and generate the hashed message PH$_M$ based on PH.

…

24: *M'* ← toByte(1, 1) || toByte(|*ctx*|, 1) || *ctx* || OID || PH$_M$

25: SIG ← slh_sign_internal(*M'*, SK, *addrnd*)

26: **return** SIG

### A. 實驗環境

由於 GSM 協會 (Groupe Speciale Mobile Association, GSMA)在 2024 年年底時發佈白皮書「IG.18 Opportunities and Challenges for Hybrid (QKD and PQC) Scenarios」，內容指出未來在核心網路設備和終端設備都有可能採用量子隨機數產生器結合後量子密碼學演算法[12]。有鑑於此，本研究主要考量終端設備的情境，在硬體的部分主要採用 Raspberry Pi 4 進行量子隨機數產生器和後量子密碼學演算法實作，如圖 3 所示。

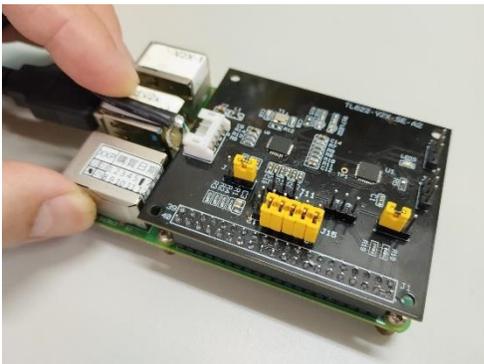

Fig. 3. 本研究實驗環境的終端設備

由於本研究主要聚焦在概念性驗證(Proof of Concept, POC)，並未實作量子晶片，所以量子計算主要建構在 IBM Qiskit SDK 1.1.1 以模擬器(simulator)方式實作。需要注意的是，未來仍需採用真實的量子晶片，才能具備量子物理特性和具備安全性。後量子密碼學演算法主要採用 BouncyCastle Open Source API 1.81 進行實作，修改內部函數來呼叫量子隨機數產生器得到需要的隨機位元值。需要注意的是，未來在此呼叫上也應該建立在安全的通道和機制上。

### B. 量子隨機數產生器的隨機位元熵驗證

本研究參考 NIST SP 800-90B 所述驗證方法[9]收集兩個資料集：

(1). Sequential dataset：產製 100 萬個隨機位元值，並且分析 0 和 1 的分布。

(2). Restart test dataset：重新啟動 1000 次，每次產製 1000 個隨機位元值，建構 1000×1000 大小的 restart 矩陣。

計算 restart 矩陣的每一行和每一列的 0 和 1 發生的次數，並且取得最高次數作為 Most Common Value (MCV)，再根據 MCV 換算為最小熵(mini_entropy)和運用二項式分布檢定得到 *p*-value。

本研究設計的 6 種量子隨機數產生器的驗證結果如表 II 和圖 4 所示。其中，由於每一行和每一列都是 1000 個隨機位元值，所以如果 0 和 1 在均勻分布下期望值為各 500 次。因此，當 MCV 越接近 500，則最小熵(mini_entropy)值越大，以及 *p*-value 也會越大。當 *p*-value 低於門檻值時，則表示 0 和 1 不是均勻分布，混亂程度太低；該門檻值在 NIST SP 800-90B [9]中定義為 0.000005。取 MCV 觀察的用意在於表示該組資料是 2000 組資料中熵值最小的那組資料，也就是 worst case；當如果 worst case 都有驗證通過，剩下的 1999 組資料也會通過。

圖 4 採用盒鬚圖的方式呈現 6 種量子隨機數產生器在各自 2000 組資料對應的 *p*-value，從分布可以觀察到全部都有高於門檻值 0.000005。表示 6 種量子隨機數產生器可以產製足夠混亂的隨機位元，通過驗證。

TABLE II.　隨機位元熵驗證結果

| QRNG | MCV | mini_entropy | *p*-value |
|---|---|---|---|
| H-Gate-Based | 555 | 0.99125 | 0.00056 |
| SX-Gate-Based | 552 | 0.99218 | 0.00112 |
| RX-Gate-Based | 559 | 0.98993 | 0.00021 |
| RY-Gate-Based | 560 | 0.98959 | 0.00017 |
| Based on P-Gate & H-Gate | 551 | 0.99248 | 0.00139 |
| U-Gate-Based | 558 | 0.99027 | 0.00027 |

### C. 量子隨機數產生器的隨機位元獨立且同分布驗證

除了驗證量子隨機數產生器可以產製足夠混亂的隨機位元外，還需驗證是否為獨立且同分布(IID)。因此，NIST SP 800-90B 定義了獨立性檢定(Independence Test, Ind. Test)、適應度檢定(Goodness-of-fit Test, GF Test)、以及最長重覆子字串長度檢定(Length of the Longest Repeated Substring Test, Length of the LRS Test)[9]，分述如下。

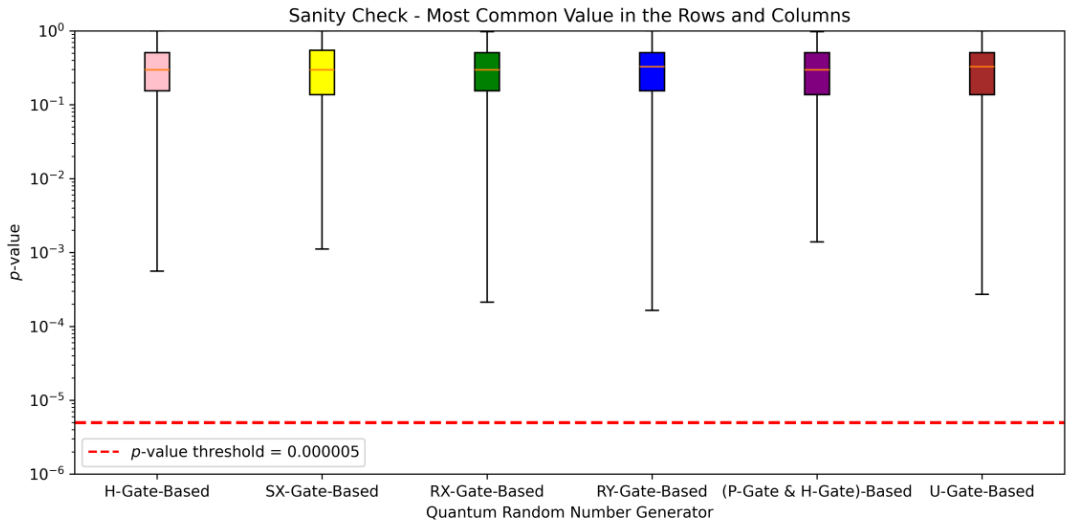

Fig. 4. Sanity check results

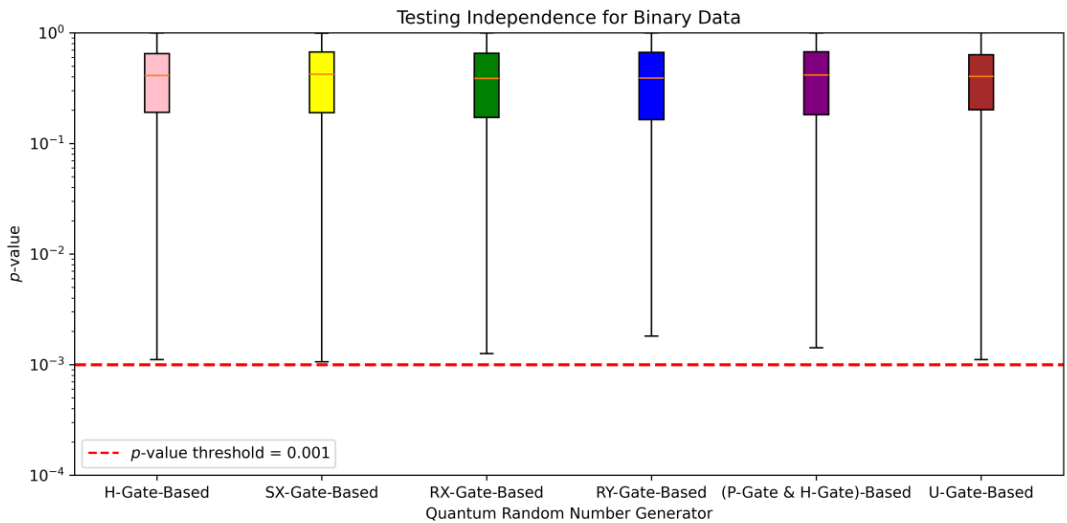

Fig. 5. Independence test results

(1). 獨立性檢定(Ind. Test)：第 VI.B 節所述 restart 矩陣每一行和每一列都是1000個隨機位元值，把每一組資料各別切割為 10 等份，每一等份裡各有 100 個隨機位元值，所以如果 0 和 1 在均勻分布下期望值為各 50 次。運用卡方檢定來驗證真值和期望值的平方差比例，並且加總後可以得每一組資料的卡方值，並且搭配自由度 9 可以計算得到 $p$-value，其中，$p$-value 門檻值在 NIST SP 800-90B [9]中定義為 0.001，對應的卡方值門檻值是 27.877。

(2). 適應度檢定(GF Test)：第 VI.B 節所述 restart 矩陣每一行和每一列都是 1000 個隨機位元值，觀察每 4 個位元的分布，把每一組資料各別切割為 250 等份，每一等份裡各有 4 個隨機位元值(即 0000~1111 的 16 種組合之一)，所以如果 0000~1111 在均勻分布下每一種組合的期望值為各 15.625 次。運用卡方檢定來驗證真值和期望值的平方差比例，並且加總後可以得每一組資料的卡方值，並且搭配自由度 14 可以計算得到 $p$-value，其中，$p$-value 門檻值在 NIST SP 800-90B [9]中定義為 0.001，對應的卡方值門檻值是 36.123。

(3). 最長重覆子字串長度檢定(Length of the LRS Test)：第 VI.B 節所述 restart 矩陣每一行和每一列都是 1000 個隨機位元值，觀察每一組資料中最長重覆子字串的長度，再運用二項式分布檢定計算 $p$-value。其中，$p$-value 門檻值在 NIST SP 800-90B [9]中定義為 0.001，所以在長度為 1000 時對應的最長重覆子字串的長度門檻值大約為 28。

本研究設計的 6 種量子隨機數產生器的隨機位元獨立且同分布驗證結果如表 III 和表 IV 所示。其中，表 III 為各組資料在各種檢定計算的 $p$-value 中位數，表 IV 為各組資料在各種檢定計算的 $p$-value 最小值(也就是 worst case)。此外，各組資料在獨立性檢定和適應度檢定的 $p$-value 以盒鬚圖呈現，分別如圖 5 和圖 6 所示。以及最長重覆子字串長

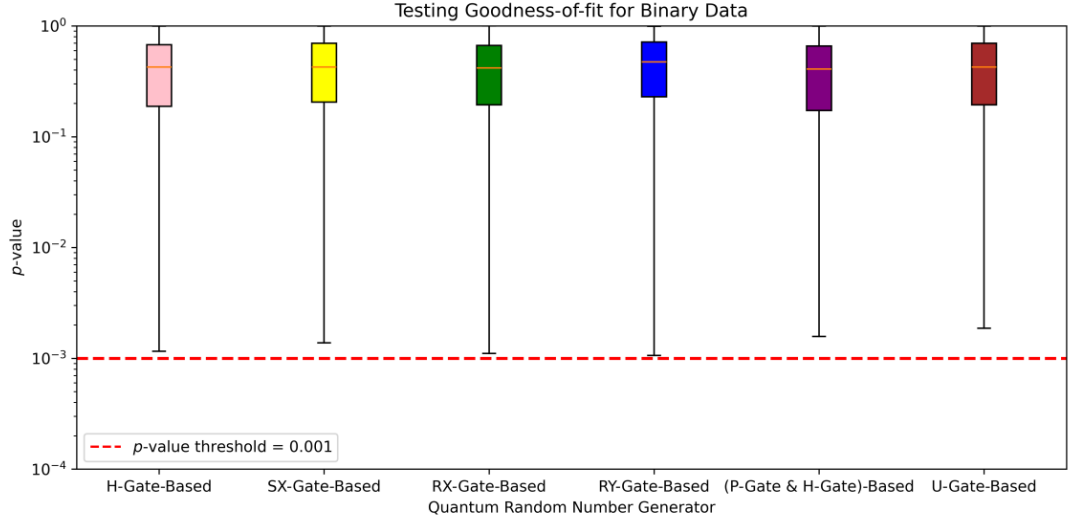

Fig. 6. Goodness-of-fit test results

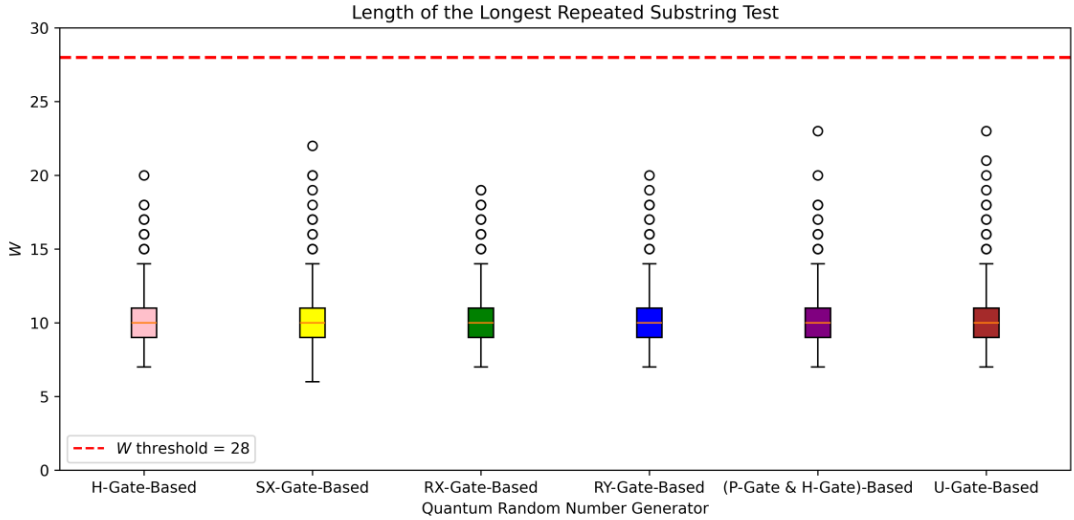

Fig. 7. Length of the longest repeated substring test results

度檢定的最長重覆子字串長度以盒鬚圖呈現，如圖 7 所示。由實驗結果可以觀察到即使在 worst case 的情況下，$p$-value 也都有高於門檻值 0.001，所以全部都有通過獨立且同分布驗證。

TABLE III. 隨機位元獨立且同分布驗證 $p$-VALUE 中位數

| QRNG | Ind. Test | GF Test | Length of the LRS Test |
|---|---|---|---|
| H-Gate-Based | 0.41184 | 0.42615 | 1.00000 |
| SX-Gate-Based | 0.42264 | 0.42615 | 1.00000 |
| RX-Gate-Based | 0.38726 | 0.41690 | 1.00000 |
| RY-Gate-Based | 0.39072 | 0.47381 | 1.00000 |
| Based on P-Gate & H-Gate | 0.41542 | 0.40774 | 1.00000 |
| U-Gate-Based | 0.40296 | 0.42615 | 1.00000 |

TABLE IV. 隨機位元獨立且同分布驗證 $p$-VALUE 最小值

| QRNG | Ind. Test | GF Test | Length of the LRS Test |
|---|---|---|---|
| H-Gate-Based | 0.00111 | 0.00116 | 0.36772 |
| SX-Gate-Based | 0.00106 | 0.00138 | 0.10787 |
| RX-Gate-Based | 0.00126 | 0.00111 | 0.60097 |
| RY-Gate-Based | 0.00181 | 0.00106 | 0.36772 |
| Based on P-Gate & H-Gate | 0.00142 | 0.00157 | 0.05536 |
| U-Gate-Based | 0.00111 | 0.00187 | 0.05536 |

D. *量子隨機數產生器計算時間比較*

本節主要在不同的(Length of Random Bits, Number of Qubits)組合下，比較本研究設計的 6 種量子隨機數產生器

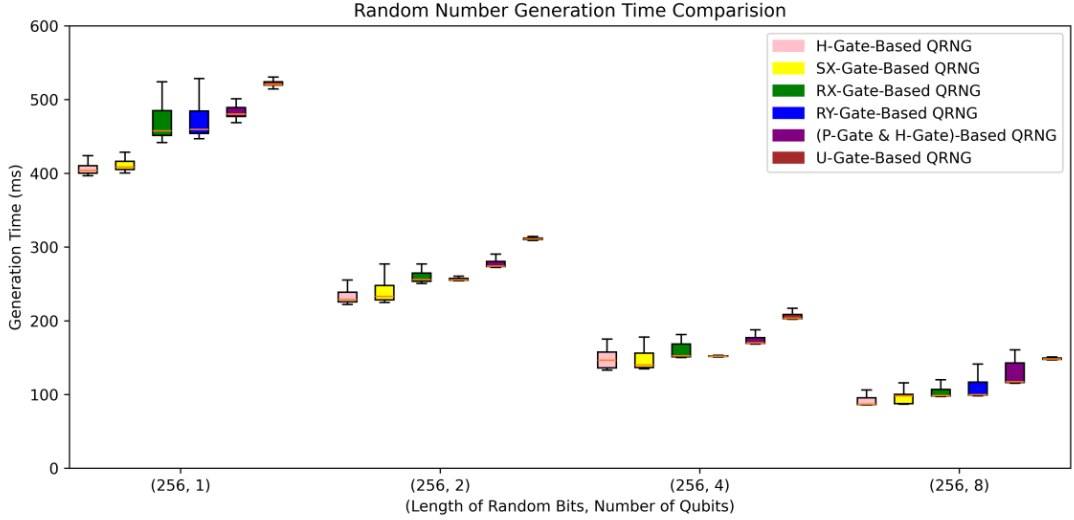

Fig. 8. Random number genertion time comparision

的計算時間。其中，由於在 ML-KEM、ML-DSA、SLH-DSA 所需要用到的隨機數長度多為 256 bits，所以實驗中主要考慮 Length of Random Bits 為 256 的情境。並且，Number of Qubits 分別考慮 1、2、4、8 的情境，當 Number of Qubits 值越小，則建構量子電路的成本越低，但相對需要循環執行較多的次數，計算時間越多；相反，當 Number of Qubits 值越大，則建構量子電路的成本越高，但相對需要循環執行較少的次數，計算時間越少。

在不同的組合各執行 1000 次，計算時間中位數如表 V 所示。為完整充分呈現實驗結果，本研究採用盒鬚圖的方式呈現，如圖 8 所示。可以觀察到(256, 1)組合的計算時間最長，而(256, 8)組合的計算時間最短。此外，未來如果能再增加 Number of Qubits 的話，則計算時間可以更短，得到更高的效率。另外，可以觀察到 H-Gate-Based QRNG 和 SX-Gate-Based QRNG 的計算上較單純，所以效率最高。而由於(P-Gate & H-Gate)-Based QRNG 和 U-Gate-Based QRNG 都較需要較多的計算量，所以計算時間最長。

TABLE V. 量子隨機數產生器計算時間比較結果(單位：毫秒)

| QRNG | (256, 1) | (256, 2) | (256, 4) | (256, 8) |
|---|---|---|---|---|
| H-Gate-Based | 403.80 | 228.83 | 146.50 | 87.00 |
| SX-Gate-Based | 408.38 | 232.95 | 139.99 | 98.34 |
| RX-Gate-Based | 457.67 | 255.79 | 152.32 | 98.45 |
| RY-Gate-Based | 459.49 | 255.43 | 151.87 | 99.84 |
| Based on P-Gate & H-Gate | 480.30 | 274.20 | 169.87 | 117.35 |
| U-Gate-Based | 520.50 | 310.82 | 203.37 | 148.19 |

### E. NIST 後量子密碼學標準演算法的計算時間比較

本節主要討論量子隨機數產生器應用於 NIST 後量子密碼學標準演算法的計算時間，依序驗證 QRNG-Based ML-KEM、QRNG-Based ML-DSA、QRNG-Based SLH-DSA 的效能。此外，本研究採用 Java 內建的 SecureRandom 類別作為 PRNG 產製隨機數，視為 benchmark 提供對比。在實驗過程中，主要各執行 1000 次，並且取實驗結果的中位數進行比較。

#### 1) The Evaluation of QRNG-Based ML-KEM

本研究驗證 QRNG-Based ML-KEM 產製金鑰對的計算時間，實驗結果如表 VI 所示。可以觀察到即使在 ML-KEM-1024 的參數組合下，benchmark 的計算時間仍在 1 毫秒內即可完成。NIST 評選和設計出來的 ML-KEM 具有很高的效率。然而，由於本研究的量子隨機數產生器主要建構在模擬器(simulator)上，所以需要較多的時間產製隨機數，並且 ML-KEM 產製金鑰對總共需要 512 bits 長度的隨機數，所以需要花費更多的時間。因此，結合本研究設計的 6 種量子隨機數產生器的 QRNG-Based ML-KEM 在產製金鑰對的計算時間主要取決於量子隨機數產生器產製 512 bits 長度隨機數所需的時間。

TABLE VI. ML-KEM 產製金鑰對計算時間比較結果(單位：毫秒)

| RNG | ML-KEM-512 | ML-KEM-768 | ML-KEM-1024 |
|---|---|---|---|
| SecureRandom PRNG (benchmark) | 0.614 | 0.872 | 0.979 |
| H-Gate-Based QRNG | 173.967 | 174.518 | 174.427 |
| SX-Gate-Based QRNG | 196.834 | 196.966 | 196.854 |
| RX-Gate-Based QRNG | 197.693 | 196.996 | 197.347 |
| RY-Gate-Based QRNG | 199.306 | 200.978 | 200.046 |
| QRNG Based on P-Gate & H-Gate | 234.547 | 235.056 | 248.954 |
| U-Gate-Based QRNG | 296.398 | 296.698 | 297.025 |

本研究亦驗證 QRNG-Based ML-KEM 產製金鑰封裝的計算時間，實驗結果如表 VII 所示。可以觀察到 benchmark 的計算時間約介於 0.413 毫秒~0.697 毫秒。然而，由於本

研究的量子隨機數產生器需要較多的時間產製隨機數，而 ML-KEM 產製金鑰封裝對總共需要 256 bits 長度的隨機數。因此，結合本研究設計的 QRNG-Based ML-KEM 在產製金鑰封裝的計算時間主要取決於量子隨機數產生器產製 256 bits 長度隨機數所需的時間。

TABLE VII. ML-KEM 產製金鑰封裝計算時間比較結果(單位：毫秒)

| RNG | ML-KEM-512 | ML-KEM-768 | ML-KEM-1024 |
|---|---|---|---|
| SecureRandom PRNG (benchmark) | 0.413 | 0.595 | 0.697 |
| H-Gate-Based QRNG | 87.131 | 87.425 | 87.390 |
| SX-Gate-Based QRNG | 98.562 | 98.648 | 98.624 |
| RX-Gate-Based QRNG | 98.983 | 98.734 | 98.853 |
| RY-Gate-Based QRNG | 99.876 | 100.599 | 100.219 |
| QRNG Based on P-Gate & H-Gate | 117.423 | 117.668 | 124.542 |
| U-Gate-Based QRNG | 148.309 | 148.506 | 148.703 |

*2) The Evaluation of QRNG-Based ML-DSA*

本研究驗證 QRNG-Based ML-DSA 產製金鑰對的計算時間，實驗結果如表 VIII 所示。可以觀察到即使在 ML-DSA-87 的參數組合下，benchmark 的計算時間約為 1.355 毫秒可以完成。NIST 評選和設計出來的 ML-DSA 具有很高的效率。然而，由於本研究的量子隨機數產生器主要建構在模擬器(simulator)上，所以需要較多的時間產製隨機數，並且 ML-DSA 產製金鑰對雖僅需 256 bits 長度的隨機數，但所需要花費的時間高於 benchmark 的計算時間。因此，結合本研究設計的 6 種量子隨機數產生器的 QRNG-Based ML-DSA 在產製金鑰對的計算時間主要取決於量子隨機數產生器產製 256 bits 長度隨機數所需的時間。

TABLE VIII. ML-DSA 產製金鑰對計算時間比較結果(單位：毫秒)

| RNG | ML-DSA-44 | ML-DSA-65 | ML-DSA-87 |
|---|---|---|---|
| SecureRandom PRNG (benchmark) | 0.905 | 1.070 | 1.355 |
| H-Gate-Based QRNG | 87.702 | 88.575 | 88.375 |
| SX-Gate-Based QRNG | 98.806 | 99.545 | 99.369 |
| RX-Gate-Based QRNG | 99.332 | 99.540 | 99.465 |
| RY-Gate-Based QRNG | 100.369 | 100.619 | 101.089 |
| QRNG Based on P-Gate & H-Gate | 117.522 | 118.806 | 117.829 |
| U-Gate-Based QRNG | 149.015 | 149.176 | 149.625 |

本研究亦驗證 QRNG-Based ML-DSA 產製數位簽章的計算時間，實驗結果如表 IX 所示。可以觀察到 benchmark 的計算時間約介於 1.390 毫秒~2.794 毫秒。由於 ML-DSA 產製數位簽章對總共需要 256 bits 長度的隨機數，所以本研究設計的 QRNG-Based ML-DSA 在產製數位簽章計算時間與產製金鑰對計算時間相似，主要取決於量子隨機數產生器產製 256 bits 長度隨機數所需的時間。

TABLE IX. ML-DSA 產製數位簽章計算時間比較結果(單位：毫秒)

| RNG | ML-DSA-44 | ML-DSA-65 | ML-DSA-87 |
|---|---|---|---|
| SecureRandom PRNG (benchmark) | 1.390 | 2.371 | 2.794 |
| H-Gate-Based QRNG | 88.665 | 91.308 | 90.826 |
| SX-Gate-Based QRNG | 99.377 | 101.092 | 100.545 |
| RX-Gate-Based QRNG | 100.276 | 101.798 | 101.653 |
| RY-Gate-Based QRNG | 101.554 | 104.424 | 104.534 |
| QRNG Based on P-Gate & H-Gate | 118.910 | 123.106 | 120.780 |
| U-Gate-Based QRNG | 149.915 | 150.914 | 151.777 |

*3) The Evaluation of QRNG-Based SLH-DSA*

由於 SLH-DSA 的參數組合較多，所以本研究主要挑選 SLH-DSA SHAKE-f 系列進行比較，本研究驗證 QRNG-Based SLH-DSA 產製金鑰對的計算時間，實驗結果如表 X 所示。可以觀察到 benchmark 的計算時間約介於 6.445 毫秒~23.560 毫秒之間。由於本研究設計的 6 種量子隨機數產生器的 QRNG-Based SLH-DSA 在產製金鑰對的計算時間主要取決於量子隨機數產生器產製隨機數所需的時間。因此，在產製 SLH-DSA SHAKE-128f 金鑰對時，僅需 384 bits 長度的隨機數，所需計算時間較短。而在產製 SLH-DSA SHAKE-192f 金鑰對和 SLH-DSA SHAKE-256f 金鑰對時，則分別需要 576 bits 長度的隨機數和 768 bits 長度的隨機數，所需計算時間較長。

TABLE X. SLH-DSA 產製金鑰對計算時間比較結果(單位：毫秒)

| RNG | SLH-DSA SHAKE-128f | SLH-DSA SHAKE-192f | SLH-DSA SHAKE-256f |
|---|---|---|---|
| SecureRandom PRNG (benchmark) | 6.445 | 9.079 | 23.560 |
| H-Gate-Based QRNG | 136.645 | 204.519 | 283.712 |
| SX-Gate-Based QRNG | 154.540 | 229.921 | 316.971 |
| RX-Gate-Based QRNG | 153.877 | 229.666 | 318.319 |
| RY-Gate-Based QRNG | 162.391 | 233.187 | 322.601 |
| QRNG Based on P-Gate & H-Gate | 182.025 | 275.916 | 375.705 |
| U-Gate-Based QRNG | 228.539 | 342.069 | 467.849 |

本研究亦驗證 QRNG-Based SLH-DSA 產製數位簽章的計算時間，實驗結果如表 XI 所示。由於 SLH-DSA 產製數位簽章所需計算時間較長，可以觀察到 benchmark 的計算時間約介於 148.651 毫秒~490.060 毫秒。但由於本研究設計的 6 種量子隨機數產生器產製隨機數的計算時間仍高於 SecureRandom PRNG 產製隨機數的計算時間，所以 QRNG-Based SLH-DSA 產製數位簽章的計算時間較長。其

中，由於在產製 SLH-DSA SHAKE-128f 數位簽章時，僅需 128 bits 長度的隨機數，所需計算時間較短。而在產製 SLH-DSA SHAKE-192f 數位簽章和 SLH-DSA SHAKE-256f 數位簽章時，則分別需要 192 bits 長度的隨機數和-256 bits 長度的隨機數，所需計算時間較長。

TABLE XI. SLH-DSA 產製數位簽章計算時間比較結果(單位：毫秒)

| RNG | SLH-DSA SHAKE-128f | SLH-DSA SHAKE-192f | SLH-DSA SHAKE-256f |
|---|---|---|---|
| SecureRandom PRNG (benchmark) | 148.651 | 234.374 | 490.060 |
| H-Gate-Based QRNG | 192.138 | 299.832 | 577.291 |
| SX-Gate-Based QRNG | 198.065 | 308.165 | 587.393 |
| RX-Gate-Based QRNG | 198.093 | 308.367 | 588.517 |
| RY-Gate-Based QRNG | 200.975 | 309.816 | 591.005 |
| QRNG Based on P-Gate & H-Gate | 208.266 | 323.529 | 608.306 |
| U-Gate-Based QRNG | 223.086 | 345.522 | 638.762 |

VII. 結論與未來研究

為進一步提升安全性，本研究結合量子隨機數產生器和 NIST 後量子密碼學標準演算法。本研究提出一個泛化(general)的量子隨機數產生器架構，並且建構 6 種量子隨機數產生器，並且設計 QRNG-Based ML-KEM、QRNG-Based ML-DSA、QRNG-Based SLH-DSA，運用量子隨機數產生器產製金鑰對、產製金鑰封裝、以及產製數位簽章。在實驗中，參考 NIST SP 800-90B 所述驗證方法[9]，驗證本研究提出的量子隨機數產生器，符合隨機位元熵(entropy)驗證條件和隨機位元獨立且同分布(IID)驗證條件。除此之外，本研究亦對計算時間進行詳盡的比較，可供未來部署參考

本研究旨在設計 QRNG-Based ML-KEM、QRNG-Based ML-DSA、QRNG-Based SLH-DSA，並且設計泛化(general)的量子隨機數產生器架構。然而，本研究在量子計實作上採用的是 IBM Qiskit SDK，並且是用模擬器(simulator)的方式，所以僅是模擬量子計算，並不具備真實量子物理特性。建議未來應該改為採用真實量子計算晶片，來提供量子物理特性，並提供真隨機數，以及通過 NIST SP 800-90B 所述驗證方法。除此之外，本研究目前僅為概念性驗證(POC)，所以是先以軟體金鑰的方式來展示，未來應該設計符合安全要求的基於量子隨機數產生器的硬體安全模組(HSM)。期望未來 NIST 對基於量子隨機數產生器的硬體安全模組能提供一些安全性指引文件或規範，感謝 NIST 卓越的貢獻。

參考文獻

[1] P. W. Shor, "Polynomial-Time Algorithms for Prime Factorization and Discrete Logarithms on a Quantum Computer," in *SIAM Journal on Computing*, vol. 26, no. 5, pp.1484-1509, 1997, doi: 10.1137/S0097539795293172.

[2] "Module-Lattice-Based Key-Encapsulation Mechanism Standard," in *Federal Information Processing Standards*, FIPS 203, pp.1-47, 13 August 2024, doi: 10.6028/NIST.FIPS.203.

[3] "Module-Lattice-Based Digital Signature Standard," in *Federal Information Processing Standards*, FIPS 204, pp.1-55, 13 August 2024, doi: 10.6028/NIST.FIPS.204.

[4] "Stateless Hash-Based Digital Signature Standard," in *Federal Information Processing Standards*, FIPS 205, pp.1-51, 13 August 2024, doi: 10.6028/NIST.FIPS.205.

[5] G. Alagic et al., "Status Report on the Third Round of the NIST Post-Quantum Cryptography Standardization Process," in *NIST Interagency/Internal Report*, NIST IR 8413-upd1, pp. 1-93, 5 July 2022, doi: 10.6028/NIST.IR.8413-upd1.

[6] G. Alagic et al., "Status Report on the Fourth Round of the NIST Post-Quantum Cryptography Standardization Process," in *NIST Interagency/Internal Report*, NIST IR 8545, pp. 1-27, March 2025, doi: 10.6028/NIST.IR.8545.

[7] G. Alagic et al., "Transition to Post-Quantum Cryptography Standards," in *NIST Interagency/Internal Report*, NIST IR 8547, pp. 1-22, 12 November 2024, doi: 10.6028/NIST.IR.8547.

[8] Y. K. Liu and D. Moody, "Post-quantum Cryptography and the Quantum Future of Cybersecurity," in *Physical Review Applied*, vol. 21, no. 4, article no. 040501, pp. 1-9, 2024, doi: 2331-7019/24/21(4)/040501(9).

[9] M. S. Turan et al., "Recommendation for the Entropy Sources Used for Random Bit Generation," in *NIST Special Publications*, NIST SP 800-90B, pp. 1-76, January 2018, doi: 10.6028/NIST.SP.800-90B.

[10] P. K. Lala, *Quantum Computing: A Beginner's Introduction*, 1st Edition, McGraw-Hill Education, New York, U.S., 2019, ISBN: 9781260123111.

[11] A. K. Ekert, "Quantum Cryptography Based on Bell's Theorem," in *Physical Review Letters*, vol. 67, no. 6, pp. 661-663, 1991, doi: 10.1103/PhysRevLett.67.661.

[12] "IG.18 Opportunities and Challenges for Hybrid (QKD and PQC) Scenarios," Whitepaper of the GSMA, Version 1.0, pp. 1-23, 20 October 2024. [Online]. Available: https://www.gsma.com/newsroom/wp-content/uploads/IG.18-Hybrid-QKD-and-PQC-security-scenarios-and-use-cases-Whitepaper-v1.0-002.pdf